\documentstyle[prl,aps,amssymb,twocolumn,psfig,latexsym]{revtex}

\begin{document}

\draft

\twocolumn[\hsize\textwidth\columnwidth\hsize\csname
@twocolumnfalse\endcsname
\title{Electron and Nuclear Spin Dynamics in Antiferromagnetic 
Molecular Rings}
\author{Florian Meier and Daniel Loss}
\address{ Department of Physics and Astronomy, University of Basel, 
Klingelbergstrasse 82, 4056 Basel, Switzerland}
\date{June 4, 2001}

\maketitle

\begin{abstract}
We study theoretically the spin dynamics of antiferromagnetic molecular rings,
such as the ferric wheel Fe$_{10}$. For a single nuclear or impurity spin 
coupled to one of the electron spins of the ring, we 
calculate nuclear and electronic spin correlation functions and show 
that nuclear magnetic resonance (NMR) and  electron 
spin resonance (ESR) techniques can be used to detect coherent tunneling 
of the N{\'e}el vector in these rings. 
The location of the NMR/ESR resonances gives the tunnel splitting and its
linewidth an upper bound on the decoherence rate of the electron spin 
dynamics. 
We illustrate the experimental feasibility of our proposal with 
estimates for Fe$_{10}$ molecules.
\end{abstract}

\vskip2pc]

Magnetic molecular clusters have been the subject of intense research in 
recent years, because they offer the possibility of observing macroscopic 
quantum phenomena~\cite{thiaville:99}. Ring systems such as the 
antiferromagnetic (AF) ferric wheels, Fe$_6$ and 
Fe$_{10}$~\cite{taft:94,caneschi:96,gatteschi:94}, also allow one to 
study the transition from microscopic magnetism to one-dimensional 
bulk magnetism. In ferromagnetic clusters such as Fe$_8$ and Mn$_{12}$, 
incoherent tunneling of the $S=10$ magnetic moment is observed directly 
in magnetization and susceptibility 
measurements~\cite{thomas:96,wernsdorfer:99}. The AF ferric wheels are 
candidates for the observation of macroscopic quantum coherence (MQC) 
in the form of coherent tunneling of the N\'eel vector~\cite{chiolero:98}. 
Although 
quantum effects in antiferromagnets are expected to be more pronounced 
than in ferromagnets~\cite{chiolero:98,barbara:90}, the detection of 
quantum behavior is experimentally more challenging . The reason for 
this is that magnetization and susceptibility measurements probe only 
the total spin of the molecule which, by symmetry, remains unaffected 
upon tunneling of the N\'eel vector ${\bf n}$. At low temperatures, the 
dynamics of ${\bf n}$ is determined by two characteristic frequencies, the
tunneling rate $\Delta/\hbar$ and the electron spin decoherence rate 
$\Gamma_S$. In Fe$_{10}$, $\Delta$ can be tuned from $0$ to $2$K by varying
the magnetic field~\cite{chiolero:98}. Although the tunnel splitting 
$\Delta$ enters thermodynamic quantities such as magnetization and 
specific heat, no conclusive results 
revealing the characteristic functional dependence $\Delta(B)$ have 
yet been obtained~\cite{affronte:99}. Measuring $\Gamma_S$ is of 
central importance for the characterization of the quantum tunneling, 
since {\it coherent} quantum tunneling requires $\Gamma_S \lesssim
\Delta$. An estimate of $\Gamma_S$ can be obtained from the
typical energy scales of various interactions leading to decoherence. 
Spin-phonon
interactions are frozen out at low $T$. Nuclear 
dipolar ($0.1$mK) and hyperfine ($1$mK) interactions are significantly 
smaller than interring electron spin dipolar interactions (some $10$mK).
These numbers indicate that, in Fe$_{10}$, tunneling of ${\bf n}$ is 
coherent for a wide range of $\Delta$.

Motivated by these numbers, in this paper we study the quantum spin dynamics 
of the ferric wheel. We show 
that the tunnel dynamics of ${\bf n}$ enters the correlation 
functions of a {\it single} electron spin. As indicated in~\cite{irkhin:86} 
for a bulk AF system, a nuclear spin coupled to a single electron spin acts 
as a local spin probe. We show that in small AF rings in which ${\bf n}$ 
itself has additional coherent dynamics, the 
correlation functions of the nuclear spin 
exhibit signatures of the tunneling of ${\bf n}$. In particular, 
we discuss the coherent dynamics 
of one nuclear spin coupled to one of the electron spins of the ferric wheel 
and show 
that both the tunnel splitting $\Delta$ and the electron spin decoherence
rate can be obtained from NMR and ESR spectra.

The ferric wheels Fe$_{6}$ and Fe$_{10}$ are well 
characterized~\cite{taft:94,caneschi:96,cornia:99,waldmann:99}. The 
$s=5/2$ Fe III ions are arranged on a ring, with an AF nearest-neighbor 
exchange coupling $J>0$ and a weak, easy-axis anisotropy directed along 
the ring axis ${\bf e}_z$. The minimal Hamiltonian for the system is
\begin{equation}
H_0 = J \sum_{i=1}^N {\bf s}_i \cdot {\bf s}_{i+1} + {\bf h} \cdot  
\sum_{i=1}^N {\bf s}_i - k_z   \sum_{i=1}^N s_{i,z}^2,
\label{eq:h}
\end{equation}
where $N=10$ or $6$ and ${\bf s}_{N+1} \equiv {\bf s}_1$, ${\bf h} = 
g \mu_B {\bf B}$, with ${\bf B}$ the external magnetic field and $g=2$ 
the electron spin $g$-factor. For Fe$_{10}$, $J=15.56 {\rm K}$ and 
$k_z = 0.0088 J$. For Fe$_6$, the values for $J$ and $k_z$ vary 
appreciably depending on the central alkali metal atom: for Na:Fe$_6$, 
$J=32.77 {\rm K}$ and $k_z = 0.0136 J$, whereas for  Li:Fe$_6$, $J = 
20.83 {\rm K}$ and $k_z = 0.0053 J$~\cite{cornia:99,normand:00}.

For $k_z=0$, the eigenstates of $H_0$ are also eigenstates of the operator 
of total spin, ${\bf S}=\sum_{i=1}^N {\bf s}_i$, with 
$E_{S,S_x}=(2J/N)S(S+1)+h S_x$~\cite{taft:94}. For $k_z \ll 2J/(Ns)^2$,
the anisotropy can be taken into account in perturbation theory in $k_z$. The
scenario changes drastically for $k_z \gtrsim 2J/(Ns)^2$, when 
${\bf n}$ (staggered magnetization) 
is localized along $\pm {\bf e}_z$. We label the states with ${\bf n}$
oriented along $\pm {\bf e}_z$ by $|\! \uparrow\rangle$ and $|\! 
\downarrow\rangle$. In a semiclassical description, 
the low-energy sector of the ferric wheel consists of two states, a ground 
state, $|g\rangle =(|\! \uparrow\rangle + | \! \downarrow\rangle )/\sqrt{2}$, 
and a first excited state, $|e\rangle =(|\! \uparrow\rangle - | \! 
\downarrow\rangle )/\sqrt{2}$. 
The static equilibrium properties of a system described by Eq.~(\ref{eq:h}) 
are discussed in detail in Ref.~\cite{chiolero:98}. With ${\bf B} 
\parallel {\bf e}_x$,  {\it i.e.} in the plane of the 
ring, the system exhibits interesting spin dynamics: In the high field regime, 
$h_x \gg \hbar \omega_0$, ${\bf n}$ is confined to the $(y,z)$-plane and 
tunneling takes place through the potential barrier of height $N k_z s^2$.
In particular, 
$\Delta= \Delta_0 |\sin (\pi N h_x/4J)|$, with  $\Delta_0 = 
8 \hbar \omega_0 \sqrt{{\mathcal S}/2 \pi\hbar} \, e^{-{\mathcal S}/\hbar}$, 
$\omega_0=s \sqrt{8 J k_z}/\hbar$, 
and  ${\mathcal S}/\hbar = N s \sqrt{2 k_z/J}$, shows oscillatory behavior 
as a function of $h_x$~\cite{chiolero:98}.
We restrict our considerations henceforth to the geometry ${\bf B} \parallel 
{\bf e}_x$. For Fe$_{10}$, $\Delta_0 \simeq 2.18$K is much larger than 
in, e.g., Mn$_{12}$~\cite{leuenberger:00,luis:00} or Fe$_{8}$. 

{\it Spin susceptibilities. -} We consider first the standard ac. spin 
susceptibility 
and ESR measurements, in which an infinitesimal magnetic probing field 
couples to the total spin ${\bf S} = \sum_{i=1}^N {\bf s}_i$ of the ferric 
wheel. We will show that these experimental techniques are insufficient 
to detect coherent tunneling of ${\bf n}$ in a system described by $H_0$ 
alone. In an effective-action description for the system (\ref{eq:h}) with 
Lagrangian density $L_E[{\bf n}]$~\cite{chiolero:98}, we find ${\bf S} = 
\frac{N}{4J} \, [i {\bf n} \times \dot{\bf n} - {\bf h} + {\bf n} ({\bf n} 
\cdot {\bf h}) ]$. In high magnetic fields $h_x \gg \hbar \omega_0$, the 
spin susceptibility, $\chi_{\alpha \alpha} (\omega) = \int_0^{\beta \hbar} 
{\rm d}\tau \, \langle T_{\tau} \hat{S}_\alpha (\tau)  \hat{S}_\alpha 
\rangle e^{i \omega_n \tau} |_{i \omega_n \rightarrow \omega + i 0 }$, 
with $\alpha = x,y,z$, $\beta =1/k_B T$, and $T_{\tau}$ the imaginary time 
ordering operator,  may be evaluated using spin path integrals. 
For $\omega, k_B T/\hbar \ll \omega_0, h_x/\hbar$, up to 
corrections ${\mathcal O}(e^{-{\mathcal S}/\hbar})$, 
\begin{equation}
\chi_{\alpha \alpha} (\omega) = \frac{N}{4J} f_{\alpha},
\label{eq:spincorrel}
\end{equation}
with $f_x = 1 - {\mathcal O} (\pi^2 N \Delta/8 J)$ and $f_y \simeq f_z
\simeq 1$.
It is clear from eq.~(\ref{eq:spincorrel}) that 
the susceptibilities $\chi_{\alpha \alpha} (\omega)$ for the ring 
(\ref{eq:h}) do {\it not} exhibit resonances at $\omega = \pm \Delta/\hbar$. 
The  main conclusion we draw from eq.~(\ref{eq:spincorrel}) is that a uniform 
magnetic field cannot drive transitions from 
$|g\rangle$ to $|e\rangle$. 
Starting from the rigid rotor Hamiltonian of the ferric 
wheel~\cite{chiolero:98}, this result can also be shown to hold for an
arbitrary direction of ${\bf B}$. 

We consider next the correlation function of a {\it single spin}, $\langle 
T_{\tau} \hat{s}_{i,\alpha} (\tau)  \hat{s}_{i,\alpha} \rangle \simeq 
s^2 \langle T_{\tau} n_\alpha (\tau)  n_\alpha \rangle$, for $h_x \ll 
4 J s$, with $i= 1, \ldots, N$. In contrast to the correlations of 
${\bf S}$ discussed above, $\langle T_{\tau} \hat{s}_{i,\alpha} (\tau) 
\hat{s}_{i,\alpha} \rangle$ indeed exhibits signatures of coherent 
tunneling of ${\bf n}$. To evaluate the correlation function, 
we use an effective two-state description for the ferric wheel and 
introduce a pseudospin $\hbar \, \mbox{\boldmath $\sigma$}/2$, with 
$|\! \uparrow\rangle$ and $|\! \downarrow\rangle$ being eigenstates 
of $\sigma_z$. The tunneling dynamics of the N{\'e}el vector ${\bf n}$ is then 
generated by the pseudo-Hamiltonian $-\Delta  \, \sigma_x/2$. Because 
$\langle T_{\tau} n_z (\tau)  n_z \rangle \simeq 
\langle T_{\tau} \sigma_z (\tau) \sigma_z  \rangle$ in the low-energy 
sector, we obtain immediately 
\begin{equation}
\langle T_{\tau} \hat{s}_{i,z} (\tau)  \hat{s}_{i,z} (\tau^\prime) 
\rangle \simeq s^2 \, \frac{\cosh[(\beta-2 |\tau - \tau^\prime|) 
\Delta/2]}{\cosh[\beta \Delta/2]}.
\label{eq:neelcorrel}
\end{equation}
After analytic continuation, the real-time correlation function exhibits 
the $e^{\pm i \Delta t/\hbar}$ time dependence characteristic of 
coherent tunneling. We conclude that {\it local} spin probes allow the 
observation of the N{\'e}el vector dynamics. Nuclear spins which 
couple (predominantly) to a given ${\bf s}_i$ are ideal candidates for
such probes, as we shall discuss next.

{\it Nuclear susceptibility. -} NMR techniques have been widely used to 
study molecular magnets~\cite{lascialfari:97etc,cornia:00}. 
In the following we show that nuclear spins can also be used as a local probe 
to detect coherent tunneling of ${\bf n}$.
For simplicity, 
we restrict our considerations to interactions of the form $H^\prime = A 
{\bf s}_1 \cdot {\bf I}$, which includes both the hyperfine contact 
interaction and the direction-independent part of the magnetic dipolar 
interaction. For $^{57}$Fe, the dominant coupling to the electron 
spin is by a hyperfine contact interaction, $A_{\rm Fe} s \simeq 
3.3$mK~\cite{garg:95}. In contrast, the interaction of 
a $^1$H nuclear spin with the electron spins is dipolar, with a 
direction-independent term $\sum_i A_i {\bf s}_i \cdot {\bf I}$. 
For AF order, the sum yields an effective coupling $A_{\rm H} s {\bf n} 
\cdot {\bf I}$, where, for Fe$_{10}$, the coefficient $A_{\rm H}=\sum_i 
(-1)^{i+1} A_i$ depends strongly on the site of the proton spin ${\bf I}$. 
For many  of the inequivalent sites, however, $A_H$ is of order 
$0.1$mK~\cite{cornia:00}. 
With $N_{\rm Fe}$ and $N_{\rm H}$ the numbers of NMR-active $^{57}$Fe 
and proton nuclei, 
as long as $N_{\rm Fe} A_{\rm Fe} + N_{\rm H} A_{\rm H} \ll \Delta$, the 
effect of the nuclear spins on the electron spin dynamics remains small. 
We define the
 decoherence rates $\Gamma_I$ and $\Gamma_S$ of the nuclear and 
electronic spins as the decay rates of $\langle I_y(t)I_y \rangle$ 
and $\langle n_z(t)n_z \rangle$, respectively. For time scales 
$t<1/\Gamma_S$, the electron spin produces a coherently oscillating 
effective magnetic field with frequency $\Delta/\hbar$ at the site of 
the nucleus. 

In order to show that this field affects the nuclear spin dynamics, we now 
consider a single, NMR-active $^{57}$Fe or $^1$H 
nucleus (inset of Fig.~1). For
$k_B T \ll  \hbar \omega_0$, we may restrict our considerations 
to the Hilbert space spanned by $\{|g\rangle,|e\rangle\} $. By using the 
decomposition ${\bf s}_i = (-1)^{i+1} s {\bf n} + {\bf S}/N$  of a single 
spin into staggered magnetization $\pm s {\bf n}$, ${\bf n}^2 = 1$,
 and fluctuations 
${\bf S} \perp {\bf n}$ around the N{\'e}el ordered state, we obtain 
$H^{\prime} = A {\bf s}_1 \cdot {\bf I} \simeq
A s {\bf n} \cdot {\bf I}$. 
We show now that, due to
$\langle e | n_z |g \rangle \neq 0$ , the tunneling dynamics of 
${\bf n}$ can be obtained from the nuclear spin correlation functions 
$\langle I_\alpha (t) I_\alpha \rangle$. With $\langle e|n_z |g 
\rangle = {\mathcal O}(1)$, the dominant term in $H^\prime$ is $A s n_z 
I_z$, or in the pseudospin notation introduced above, 
\begin{equation}
H^\prime \simeq A s n_z I_z \simeq A s \sigma_z  I_z.
\label{eq:h-si2}
\end{equation}
NMR experiments measure via power absorption the 
imaginary part of the nuclear spin susceptibility, $\chi^{\prime 
\prime}_{I,\alpha \alpha}(\omega)$, and by pulsed techniques the 
nuclear spin correlation functions $ \langle I_\alpha (t) I_\alpha 
\rangle$ in the time domain~\cite{slichter}. From expanding $\langle 
I_\alpha (t) I_\alpha \rangle$ in $H^\prime$, 
\begin{eqnarray}
&& \langle I_\alpha (t) I_\alpha \rangle \simeq \langle I_\alpha (t) 
I_\alpha  \rangle_0  - A^2 s^2/\hbar^2  \int^t_{-\infty} {\rm d}t^\prime 
\, \int^{0}_{-\infty} {\rm d}t^{\prime \prime}  \label{eq:i-correl} \\
&& \hspace*{1cm} \times \langle [I_z(t^{\prime}), I_\alpha(t)] 
[I_z(t^{\prime \prime}), I_\alpha] \rangle_0 \langle n_z(t^\prime) 
n_z (t^{\prime \prime}) \rangle_0, 
\nonumber
\end{eqnarray}
it is evident that the dynamics of ${\bf n}$ enters $\chi^{\prime 
\prime}_{I,\alpha \alpha}(\omega)$. 

To evaluate eq.~(\ref{eq:i-correl}), we diagonalize the Hamiltonian
\begin{equation}
H = - \frac{\Delta}{2}  \, \sigma_x + \gamma_I B_x I_x +  A s  I_z  \sigma_z,
\label{eq:h-si3}
\end{equation}
with $\gamma_I$ the nuclear gyromagnetic ratio, which describes the ferric 
wheel in the low-energy sector with a single 
nuclear spin ${\bf I}$ coupled to ${\bf s}_1$, eq. (\ref{eq:h-si2}). 
We assume thermal equilibrium 
for both electron and nuclear spins. For Fe$_{10}$ in the high field regime, 
the results may be derived by expansion to leading order in $A s/\Delta$ and 
$\gamma_I B_x/\Delta$, because $\Delta_0 \gg \gamma_I B_x, A s$, for both 
$^{57}$Fe and $^1$H nuclei, and $B_x \simeq 10 {\rm T}$. For $I=1/2$, 
$\chi^{\prime \prime}_{I,zz}(\omega)$ displays the unperturbed emission and 
absorption 
peaks at $\omega \simeq \pm \gamma_I B_x/\hbar$, although these are 
slightly shifted if the hyperfine term $A {\bf S}\cdot {\bf I}/N$ is 
taken into account. $\chi^{\prime \prime}_{I,xx}(\omega)$, however, displays 
resonant absorption and emission of 
small intensity at $\omega = \pm (\Delta \pm \gamma_I B_x)/\hbar$. Finally,
\begin{eqnarray}
&&\chi^{\prime \prime}_{I,yy}(\omega)= \frac{\pi}{4} \Bigl[ \tanh \left( 
\frac{\beta \gamma_I B_x}{2}\right) \, \delta(\omega-\gamma_I B_x/\hbar) 
\label{eq:i-susc}  \\
&& + \left( \frac{A s}{\Delta} \right)^2 \! \tanh \left( 
\frac{\beta \Delta}{2} \right) \delta(\omega-\Delta/\hbar) \Bigr]
- [\omega \rightarrow - \omega] 
  \nonumber
\end{eqnarray}
exhibits {\it satellite resonances at the tunnel splitting $\Delta$ of the 
electron spin system.} Their physical origin is readily understood in terms
of a classical vector model. For $A=0$, ${\bf I}(t)$
precesses around the static magnetic field ${\bf B}=B_x {\bf e}_x$. 
For $A\neq 0$, the coherent tunneling of ${\bf n}$ leads to an additional 
ac hyperfine 
field $A {\bf s}_1 (t) \simeq A s \cos (\Delta t/\hbar) {\bf e}_z$ 
at the site of the nucleus. 
In contrast to a static hyperfine field which
induces a change in precession frequency and axis, 
the rapidly oscillating hyperfine field in Fe$_{10}$ leads only to
a small deviation $\delta {\bf I}(t)={\mathcal O}(As/\Delta)$ from the 
original precession.
In particular, $\delta I_y(t) \propto 
(A s/\Delta) \sin (\Delta t/\hbar)$ also oscillates at frequency
$\Delta/\hbar$ and hence gives rise to the second term in 
eq.~(\ref{eq:i-susc}).

We restricted the above analysis to the low-energy sector of the ferric 
wheel. To check this approximation, we have performed exact 
numerical diagonalization (ED) on a small AF ring with one nuclear spin 
of $I=1/2$ coupled to one of the electron spins. For the small systems 
accessible by ED, in this case $N=4$, $s=3/2$, and $k_z=0.2 J$, the 
field range for which the theoretical framework is applicable becomes 
rather small: $2J \ll h_x \ll 6J$. However, the numerical results (Fig.~1) 
for $\langle e | I_y | g\rangle$ indicate 
that our analytical value $|\langle e | I_y | g\rangle| = A s/2 \Delta$ 
entering 
eq.~(\ref{eq:i-susc}) is a good approximation. For our parameters, the 
analytical value tends to overestimate $\langle e | I_y | g\rangle$ by 
$\sim 30$\% which results from the neglect of $n_y$ in 
eq.~(\ref{eq:h-si2}). 

We turn next to a discussion of the experimental feasibility of measuring 
$\Delta$ by NMR. Because $A s/\Delta \ll 1$, the intensity of 
the satellite peaks at $\omega = \pm \Delta/\hbar$ (\ref{eq:i-susc}) 
is small compared to that of the main peaks at $\omega = \pm \gamma_I 
B_x/\hbar$. However, this satellite peak intensity may be increased 
significantly by tuning $B_x$ close to one of the critical values 
$B_x^c$ at which the magnetization of the ferric wheel jumps and 
$\Delta(B_x=B_x^c)=0$. Note, however, that our theory only applies to high 
magnetic fields, $B_x > 7.7$T for Fe$_{10}$.
Coherent tunneling of the N{\'e}el vector ${\bf n}$ 
requires $\Delta 
\gtrsim \Gamma_S$. From the estimates for the decoherence rate $\Gamma_S$ 
given in the introduction we conclude that this condition can be satisfied 
even for a large range of $\Delta/\Delta_0 \ll 1$.

We consider first $^{57}$Fe, 
with $\gamma_I = 0.18 \mu_N$~\cite{evans:95}. For $T \simeq 2 $K and 
$B_x \sim 10$T, the relative intensity of the satellite peak at $\Delta 
= \Delta_0=2.18$K, is $(A s/\Delta)^2 \tanh (\beta \Delta/2)/  \tanh 
(\beta \gamma_I B_x/2) \simeq 0.007$. This intensity, however, increases 
by a factor of 10 (100) for $\Delta = 0.1 \Delta_0$ ($\Delta = 0.01 
\Delta_0$). For $^{1}$H with $\gamma_I = 5.59 \mu_N$~\cite{evans:95}, 
and a typical value $A s\simeq 0.1 $mK, the relative peak intensity is 
$2.05 \times 10^{-7}$ ($\Delta = \Delta_0$), $2.25 \times 10^{-6}$ 
($\Delta = 0.1 \Delta_0$), and $2.25 \times 10^{-5}$ ($\Delta = 0.01 
\Delta_0$). However, the number of protons in the ring is much larger 
than that of NMR-active $^{57}$Fe nuclei, $10 \lesssim N_{\rm H}/N_{\rm 
Fe} \lesssim 100$, depending on the doping with $^{57}$Fe. Taking into 
account that the sensitivity of proton NMR is larger than that of Fe NMR 
by a factor of $3\times 10^{4}$~\cite{evans:95}, $^{57}$Fe and proton NMR 
appear to be similarly appropriate for detecting the coherent tunneling 
of ${\bf n}$. The observation of the satellite peak in 
(\ref{eq:i-susc}) is feasible, but still remains a 
challenging experimental task. The experiment must be conducted with 
single crystals of Fe$_{10}$ (or a Fe$_6$ system with sufficiently large 
$k_z > 2 J/(N s)^2$) at high, tunable  fields ($10$T) and low 
temperatures ($2$K). 
Moreover, because $B_x^c$ depends sensitively on the relative orientation 
of ${\bf B}$ and the easy axis~\cite{cornia:99,normand:00}, careful field 
sweeps are necessary to ensure that $\Delta/\Delta_0 \ll 1$ is 
maintained~\cite{lascialfari:disc}. Note that the NMR experiment 
suggested here could 
be more easily realized with  nuclear spins exhibiting higher NMR 
sensitivity than $^{57}$Fe.

We show now that, from NMR spectra, also an upper
bound for $\Gamma_S$ can be extracted. 
The NMR resonance lines are broadened by the decoherence of the 
nuclear spin, with width $\Gamma_I$ at $\pm \gamma_I B_x/\hbar$. The 
NMR resonances at $\omega = \pm\Delta/\hbar$ also involve correlation 
functions of ${\bf n}$, see (\ref{eq:i-correl}). Thus the decoherence 
of the electron spin, $\Gamma_S$, adds to the linewidth, and the width 
of the satellite peak, $\delta$, is bounded by $\Gamma_I + \Gamma_S < 
\delta$. Measurement of $\delta$ then provides an upper bound for 
$\Gamma_S$. Further, $\Gamma_S \simeq \Delta$ marks the 
transition  from coherent to incoherent tunneling dynamics. 
Hence, if $\delta < \Delta$ this
would indicate unambigously that quantum tunneling of ${\bf n}$ is coherent. 
Note that the maximum peak height of a Lorentzian resonance line is 
${\mathcal O}(1/\delta)$, so a large $\Gamma_S$ ($< \Delta$) would 
make detection of the satellite peak increasingly difficult.

{\it ESR measurements in the presence of hyperfine interaction. -}
We show now that, in the presence of a hyperfine coupling $H^\prime$, 
the electron spin 
susceptibility of the ring, $\chi_{\alpha \alpha} (\omega)$, also exhibits the 
signature of a coherent tunneling of ${\bf n}$. This results from the fact 
that integration over the initial and final nuclear spin configurations 
causes the matrix elements $\langle e | {\bf S}|g \rangle$ occurring in the 
spectral representation of $\chi_{\alpha \alpha} (\omega)$ 
(eq. (\ref{eq:spincorrel})) to become finite. 
In the high field 
limit $h_x \gg \hbar \omega_0$, the matrix elements 
become $|\langle e | S_y | g \rangle| \simeq  (A s/2 h_x) (\Delta_0 
{\mathcal S}/4 N k_z s^2) |\cos (\pi N h_x/4J)|$ and $|\langle e | S_z 
| g \rangle| \simeq A s/2 h_x$. For a small system ($N=4$, $s=3/2$), we 
have again confirmed the qualitative features of these results by ED.
It follows that $\chi^{\prime \prime}_{\alpha \alpha} (\omega \sim 
\Delta/\hbar) = \pi |\langle e | S_\alpha | g \rangle|^2 \, \tanh (\beta 
\Delta/2) \delta (\omega - \Delta/\hbar)$ exhibits resonances at 
$\Delta/\hbar$. A qualitative understanding of this result may be 
obtained by noting that a nuclear spin polarized along ${\bf B}$ 
results in an effective magnetic field $ {\bf e}_x A/2 g \mu_B$ acting only 
on ${\bf s}_1$. In a classical description this hyperfine field causes
${\bf S}$ to acquire a component $ {\bf n} \, A s/2 h_x $ along ${\bf n}$, 
and the coherent tunneling of ${\bf n}$ now also results in an 
oscillation of the total spin ${\bf S}$. Again, the decoherence rate 
of these oscillations, and hence the linewidth of the ESR resonance, 
is bounded by $\Gamma_S$. Note that $As/2 h_x=1.2 \times 10^{-4}$ for 
a single $^{57}$Fe nucleus with $B_x=10$T, so the ESR signal is very 
weak in this case. 

However, our calculations apply to any impurity spin 
${\bf j}$, which interacts with a single electron spin only, $H^\prime 
= A {\bf s}_1 \cdot {\bf j}$. In particular, for an electronic
${\bf j}$, $A$ is typically $10^3$ times larger than for a nuclear spin, 
and ESR techniques become a valuable tool for detecting the 
tunneling of ${\bf n}$. One advantage of this technique is that it is 
no longer necessary to have $\Delta/\Delta_0 \ll 1$ to obtain a large 
signal intensity, and thus the complete range of tunnel frequencies 
could be explored experimentally~\cite{remark}. 
Our calculations also apply to situations
in which several impurity spins ${\bf j}_i$ produce different net magnetic
fields for the two sublattices of the AF ring. For illustration, we discuss 
two simple scenarios. We consider $N/2$ impurity 
spins ${\bf j}_i$ ($j_i=1/2$) coupled to ${\bf s}_1$, ${\bf s}_3$,
\ldots, $H^\prime = A \sum_{i=1}^{N/2} {\bf j}_{2i-1} \cdot 
{\bf s}_{2i-1}$. For $h_x \gg As, k_B T$, all ${\bf j}_i$ align with the 
magnetic field ${\bf B}$. Since they all couple to one sublattice only,
their net magnetic fields acting on ${\bf s}_i$ add up, 
$|\langle e|S_z|g\rangle|\simeq (N/2)As/2h_x$, leading to a $(N/2)^2$-fold 
enhancement of the ESR-signal of a single impurity. In contrast, a single 
impurity ${\bf j}$ coupled to both ${\bf s}_1$ and ${\bf s}_2$,
$H^\prime = A {\bf j} \cdot ({\bf s}_1+ {\bf s}_2)$, results in the same
net magnetic field acting on both sublattices, $\langle e|S_z|g\rangle=0$.

{\it Conclusions -} We have shown that NMR and ESR 
techniques can be used to measure both the tunnel splitting $\Delta$ 
and the decoherence rate $\Gamma_S$ in the ferric wheel. For Fe$_{10}$, we 
showed that our proposal is within experimental reach. 
Our considerations apply to any AF ring system described by $H_0$, 
eq. (\ref{eq:h}), with some impurity spin coupled to one of the electron 
spins. Hence, the proposed scheme may prove useful for
a wide class of molecular magnets. 

This work was supported by the European Network 
MolNanoMag, grant number HPRN-CT-1999-00012, the BBW Bern, and the
Swiss NSF. We are grateful to A.~Cornia,
A.~Lascialfari, M.~Leuenberger, S.~Meier, and B.~Normand for stimulating 
discussions. We are particularly indebted to A. Lascialfari for useful
discussions on the experimental realization.

\begin{figure}[f]
\centerline{\psfig{file=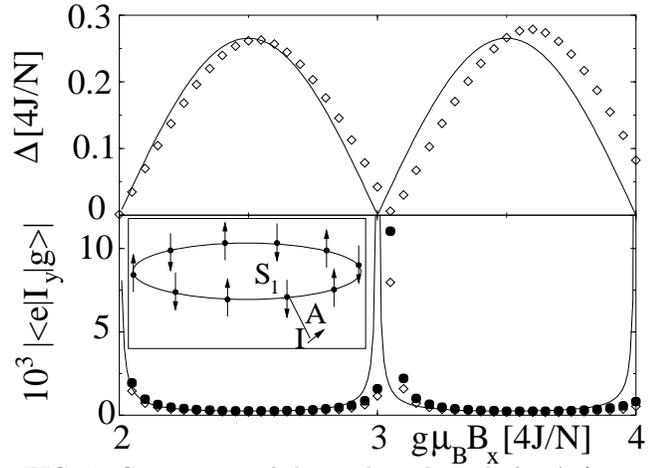,width=8.5cm}}
\caption{Comparison of the analytical result for $\Delta$ (upper panel) 
and $\langle e | I_y | g\rangle$ (lower panel) with exact numerical results 
for a small system, $N=4$, $s=3/2$, $k_z = 0.2 J$, and $A=9 \times 10^{-5}J$. 
The numerical values $(\diamond)$ for $\langle e | I_y | g\rangle$ are well 
approximated by $A s/2 \Delta$ (solid line), with a small offset in $B_x$ 
resulting from the shift of the magnetization steps when $k_z \neq 0$. For 
reference, the ratios $A s/2\Delta$ with the numerical values for $\Delta$ 
are shown ($\bullet$) in the lower panel . }
\end{figure} 


\begin{references}
\bibitem{remark}Note that the experimentally accessible quantity is the 
susceptibility
of the total spin ${\bf S}+{\bf j}$, with $\chi^{\prime \prime}_{tot,\alpha
\alpha} (\omega \sim \Delta/\hbar) = \pi 
|\langle e | S_\alpha +j_\alpha| g \rangle|^2 \, \tanh (\beta \Delta/2) 
\delta (\omega - \Delta/\hbar)$. As longs as $As\ll \Delta$,
$\langle e | j_\alpha| g \rangle$  can be evaluated as for nuclear spins.
Due to the different dependence on $h_x$, the two contributions arising from 
${\bf S}$ and ${\bf j}$ can be easily distinguished.
\end{references}
\end{document}